\documentclass[runningheads,orivec]{llncs}

\usepackage[T1]{fontenc}
\usepackage{graphicx}
\usepackage{amsmath}
\usepackage{hyperref}
\usepackage{booktabs}
\usepackage{multirow}
\usepackage{todonotes}
\usepackage{tikz}
\usetikzlibrary{positioning,arrows.meta,shapes.geometric}

\usepackage{amsmath}
\usepackage{amssymb}
\usepackage{bbm}
\usepackage{dsfont}
\usepackage{amsmath}
\usepackage[bb=dsserif]{mathalpha}
\usepackage{bm}

\usepackage[most]{tcolorbox}
\usepackage{courier} 

\newcounter{promptboxcounter}
\renewcommand{\thepromptboxcounter}{Prompt~\arabic{promptboxcounter}}

\newtcolorbox[auto counter]{promptbox}[2][]{%
  colback=gray!5!white,
  colframe=gray!75!black,
  fontupper=\ttfamily\footnotesize,
  fonttitle=\bfseries\footnotesize,
  title={\thepromptboxcounter: #2},
  label=#1,
  enhanced,
  breakable
}

\begin{document}

\title{IoT Miner -- Intelligent Extraction of Event Logs from Sensor Data for Process Mining}
\titlerunning{IoT Miner -- Intelligent Extraction of Event Logs from Sensor Data}
\author{Edyta Brzychczy \inst{1} \orcidID{0000-0002-0315-5636}  \and Urszula Jessen \inst{2,3} \orcidID{0000-0002-7282-8451} 
\and Krzysztof Kluza \inst{1} \orcidID{0000-0003-1876-9603} \and Sridhar Sriram \inst{2} \and Manuel Vargas Nettelnstroth\inst{4}}
\authorrunning{Edyta Brzychczy et al.}

\institute{%
AGH University of Krakow, Mickiewicza Av. 30, 30-059 Krakow, Poland \\ \email{\{brzych3,kluza\}@agh.edu.pl}
\and
process.science GmbH \& Co. KG, Finkenau 1, 22081, Hamburg, Germany \email{\{uj,sridhar.sriram\}@process-science.com}
\and
Eindhoven University of Technology, Eindhoven, The Netherlands \email{u.a.jessen@tue.nl}
\and
talpasolutions GmbH, Bismarckstraße 57, 45128 Essen, Germany \email{manuel@talpa-solutions.com}
}

\maketitle

\begin{abstract}
This paper presents \textit{IoT Miner}, a novel framework for automatically creating high-level event logs from raw industrial sensor data to support process mining. In many real-world settings, such as mining or manufacturing, standard event logs are unavailable, and sensor data lacks the structure and semantics needed for analysis. IoT Miner addresses this gap using a four-stage pipeline: data preprocessing, unsupervised clustering, large language model (LLM)-based labeling, and event log construction.
A key innovation is the use of LLMs to generate meaningful activity labels from cluster statistics, guided by domain-specific prompts. We evaluate the approach on sensor data from a Load-Haul-Dump (LHD) mining machine and introduce a new metric, Similarity-Weighted Accuracy, to assess labeling quality. Results show that richer prompts lead to more accurate and consistent labels. 
By combining AI with domain-aware data processing, IoT Miner offers a scalable and interpretable method for generating event logs from IoT data, enabling process mining in settings where traditional logs are missing.

\end{abstract}

\section{Introduction}
Industrial processes generate vast volumes of low-level data as a result of machine operations and monitoring equipment. While these low-level data, such as motor currents, location data, or velocity readings, offer detailed insights into process dynamics, they lack semantic clarity. 
Sensor data in its raw form is unsuitable for standard process mining (PM) techniques, which require well-defined, high-level event logs to develop process models and provide in-depth process analysis~\cite{vanderAalst2022,Weerdt_PMhandbook}. 

Bridging the semantic gap between raw sensor readings and high-level event labels remains one of the main challenges in the Business Process Management (BPM) domain \cite{Janiesh,mangler2024internet}.
To bridge this gap, a preprocessing step known as event abstraction is typically applied, which involves mapping fine-grained data into meaningful process activities \cite{Koschmider2019,Zelst}. 

Event abstraction is a challenging task, often reliant on expert rules (as a~supervised task) or data clustering (as an unsupervised task) \cite{Diba2020}. Moreover, traditional methods for event abstraction are time-consuming and difficult to scale across different processes, requiring domain knowledge to create rules or label the discovered clusters \cite{bertrand2022}.
This challenge is particularly evident in industrial contexts, where data overload and lack of expert knowledge hinder effective process analysis. Recent case studies show how integrating AI techniques, including LLMs and process mining, can help non-experts interact with complex manufacturing data and obtain rapid insights~\cite{jessen2024data}.

To overcome the mentioned disadvantages in event abstraction, we propose an IoT Miner, solution for unsupervised analysis of raw sensor data and event abstraction with the use of LLMs.
IoT Miner enables conversion, step by step, of low-level event data into high-level event logs. For this purpose, we use the following analytical pipeline: (1) raw data preprocessing, (2) clustering, (3)~labeling, and finally, (4) event logs creation.

In recent years, LLMs have demonstrated high abilities in language processing, pattern recognition, and semantic labeling, also for BPM and PM  domains~\cite{Berti_2023,brennig2025revealing,Grohs2024,vidgof_2023}. Their potential to interpret textual and structured input and adapt to new domains suggests a promising direction for the abstraction task \cite{FaniSani2024,Rebmann2025,Shirali2025}. Our first attempt to use LLM for raw sensor data labeling 
using automatically generated rules for event abstraction
was presented in \cite{Brzychczy2025}. However, in this approach, the labels for process activities were a prerequisite for LLM.

Hence, in the next step, in IoT Miner, we designed a pipeline in which the LLM automatically generates meaningful labels for the discovered clusters.
Several prompt versions are evaluated, progressively enriched with contextual information to assess how contextual information influences label quality. In evaluation, we use the real-life Load-Haul-Dump (LHD) machine dataset with ground truth data as a benchmark for evaluating the semantic similarity and consistency of the generated labels.

The paper is structured as follows: Section 2 presents related work regarding event abstraction of low-level data and LLMs usage. Section 3 describes in detail the IoT Miner solution and its architecture. Section 4 contains evaluation with an experimental setup, results, and discussion. Section 5 concludes the paper and highlights future work.

\section{Related Work}
The evolution of IoT research has progressed from basic sensing in constrained environments towards intelligent systems capable of real-time analytics. Initial IoT applications, rooted in Wireless Sensor Networks (WSNs), emphasized energy-efficient data aggregation and simple statistical methods~\cite{krishnamurthi2020overview,yao2003query}. With increased sensor deployments, cloud computing facilitated classical machine learning (ML) and time-series analytics~\cite{mahdavinejad2018machine,suryavansh2021data}, and subsequent innovations like real-time stream processing and fog computing addressed latency demands~\cite{sasaki2021survey}. Recent advances including TinyML, federated learning, and hybrid AI approaches integrate local inference with semantic reasoning, forming a strong foundation for connecting IoT with business process analytics~\cite{An_Zhou_Zou_Yang_2024,krishnamurthi2020overview}. However, a significant gap persists: methods to semantically abstract and align IoT-generated data with high-level business processes remain underdeveloped, limiting operational and organizational actionability.

In process mining, structured event logs typically originate from ERP systems or workflow engines~\cite{vanderAalst2022}, but IoT introduces opportunities to enhance this analysis with detailed sensor observations, bridging physical and digital domains~\cite{brzychczy2025process}. Despite this potential, IoT data's volume, heterogeneity, and lack of structure complicate integration into traditional process mining methods~\cite{brzychczy2025process}. To address this, frameworks like the three-phase model of Diba et al.~\cite{Diba2020} and Bertrand et al.'s conceptual UML-based model have emerged~\cite{bertrand2022}. Recent studies further enhance abstraction and real-time analysis through methods by Koschmider et al.~\cite{Koschmider_Janssen_Mannhardt}, Seiger et al.~\cite{Seiger_Franceschetti_Weber_2023}, and others~\cite{bertrand2022,Valderas_Torres_Serral_2023,Brzychczy2025}. Nonetheless, the challenge of creating generalizable, scalable, and semantically enriched event logs from IoT sensor data persists, often limited by reliance on static rules or domain-specific heuristics.

In this work, we introduce a novel approach to bridge the semantic gap between raw IoT outputs and business process constructs, with particular attention to context modeling and event granularity. The following sections detail the foundation, methodological design, and evaluation of our proposed technique.

\section{The IoT Miner Approach}

To overcome this difficulty, we present a hybrid, modular method for semantic event abstraction that combines large language models (LLMs), unsupervised clustering, and statistical preprocessing. With the help of this technique, noisy IoT signals can be automatically converted into semantically meaningful event logs, which makes it easier to use them directly for process mining and analysis. 

\subsection{System Pipeline}

\begin{figure}[ht]
\centering
\includegraphics[width=\textwidth]{  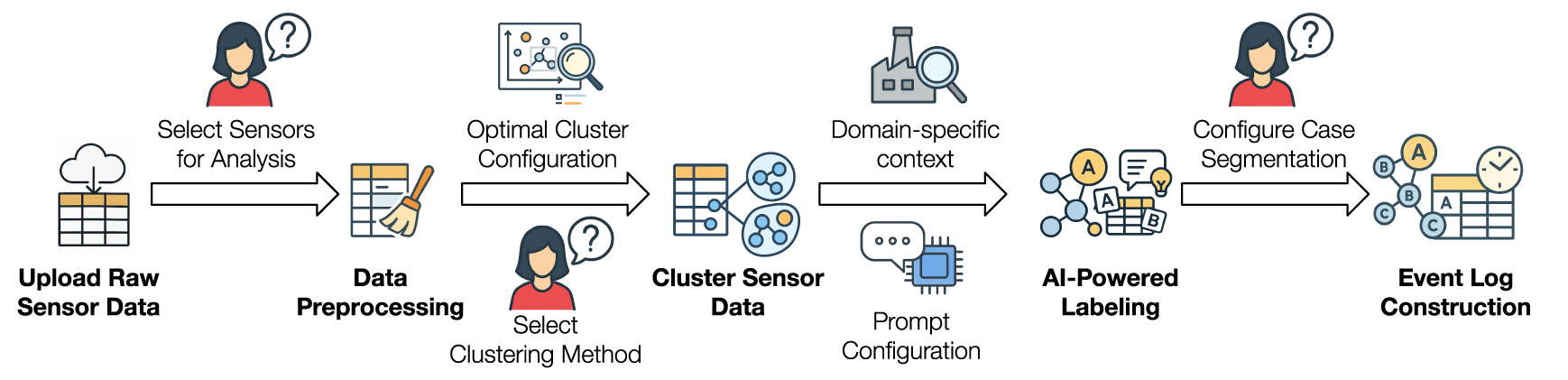}
\caption{IoT Miner pipeline}
\label{fig:pipeline}
\end{figure}

IoT Miner implements an analytical pipeline transforming raw industrial sensor data into an event log suitable for process mining purposes (Fig.~\ref{fig:pipeline}). The system connects high-frequency IoT readings with activity-level abstractions required for workflow discovery and analysis.

The pipeline integrates: (1) domain-aware preprocessing and feature engineering, (2) comparative clustering for activity discovery, (3) LLM-powered semantic labeling, and (4) intelligent event log construction with automated case segmentation. Each stage incorporates industrial domain knowledge while maintaining cross-context adaptability.

\subsection{Data Preprocessing} 
Industrial sensor data presents various challenges for process mining due to its raw, unstructured nature~\cite{bertrand2022}. IoT Miner implements a comprehensive preprocessing pipeline to transform this data into a format suitable for meaningful pattern extraction and activity identification. 
Our preprocessing approach addresses three key challenges: (1) heterogeneous data formats and quality issues, (2) high dimensionality with numerous irrelevant signals, and (3) the need to capture temporal patterns that indicate process activities. The preprocessing pipeline consists of several designed stages: 
\begin{enumerate}    
\item Data cleaning and integration --- we implement automatic detection of data types, delimiters, and encoding formats to support various input sources (CSV, Excel, JSON). For missing values, context-aware interpolation is applied based on the sensor type and operational characteristics,
\item Temporal feature engineering --- we transform the data to better reflect state changes by converting continuous sensor values to binary movement indicators using differential coding, computing first and second-order derivatives to capture rate of change information, and introducing sliding window aggregations to detect sustained patterns versus transient noise,
\item Sensor selection --- the system automatically identifies sensor columns by excluding common metadata fields (timestamps, IDs) and applying statistical measures to detect meaningful variation. 
\end{enumerate} 
Another key step is normalization, which ensures that different signals are on comparable scales—important for effective clustering. IoT Miner supports several normalization methods, including standard scaling (mean-zero, unit variance), min-max normalization, and more robust techniques based on medians and quartiles. These options help adapt the pipeline to various types of sensor data, especially in the presence of noise or outliers.

The preprocessing module is made to be both automated and adaptable. The~pipeline can be adjusted by domain experts to fit particular machines, industries, or data properties, but default configurations often function well. Because of its adaptability, IoT Miner can be used in a variety of sensor environments while still staying focused on the requirements of actual industrial processes.

\subsection{Clustering and Analysis} 
After preprocessing, IoT Miner uses unsupervised learning techniques to identify meaningful patterns in the sensor data that correspond to operational activities. The key challenge is determining the appropriate abstraction level—too fine-grained clustering captures noise rather than activities, while too coarse-grained clustering misses important operational distinctions. To address this challenge, we implement a comprehensive comparative clustering approach that automatically evaluates multiple configurations across different algorithms and parameters to identify the optimal clustering strategy for a given dataset. This approach eliminates the need for manual tuning of clustering parameters, which is typically a bottleneck in industrial applications. 

The system integrates two well-established clustering algorithms—K-means and DBSCAN—each offering distinct strengths for analyzing industrial sensor data~\cite{krishnamurthi2020overview}. 
For each algorithm, IoT Miner explores a wide range of configurations by varying both normalization methods and algorithm-specific parameters such as distance metrics or density thresholds. 
Each configuration is evaluated using standard internal validation metrics, including the Silhouette Score, Davies-Bouldin Index, and Calinski-Harabasz Index. These measures help identify the most coherent and well-separated cluster structures, laying a strong foundation for the subsequent semantic labeling of operational activities.
The system automatically selects the best algorithm for each dataset, as optimal clustering approaches may vary depending on the nature of the industrial process. 

\begin{figure}[ht]
\centering
\includegraphics[width=0.76\textwidth]{  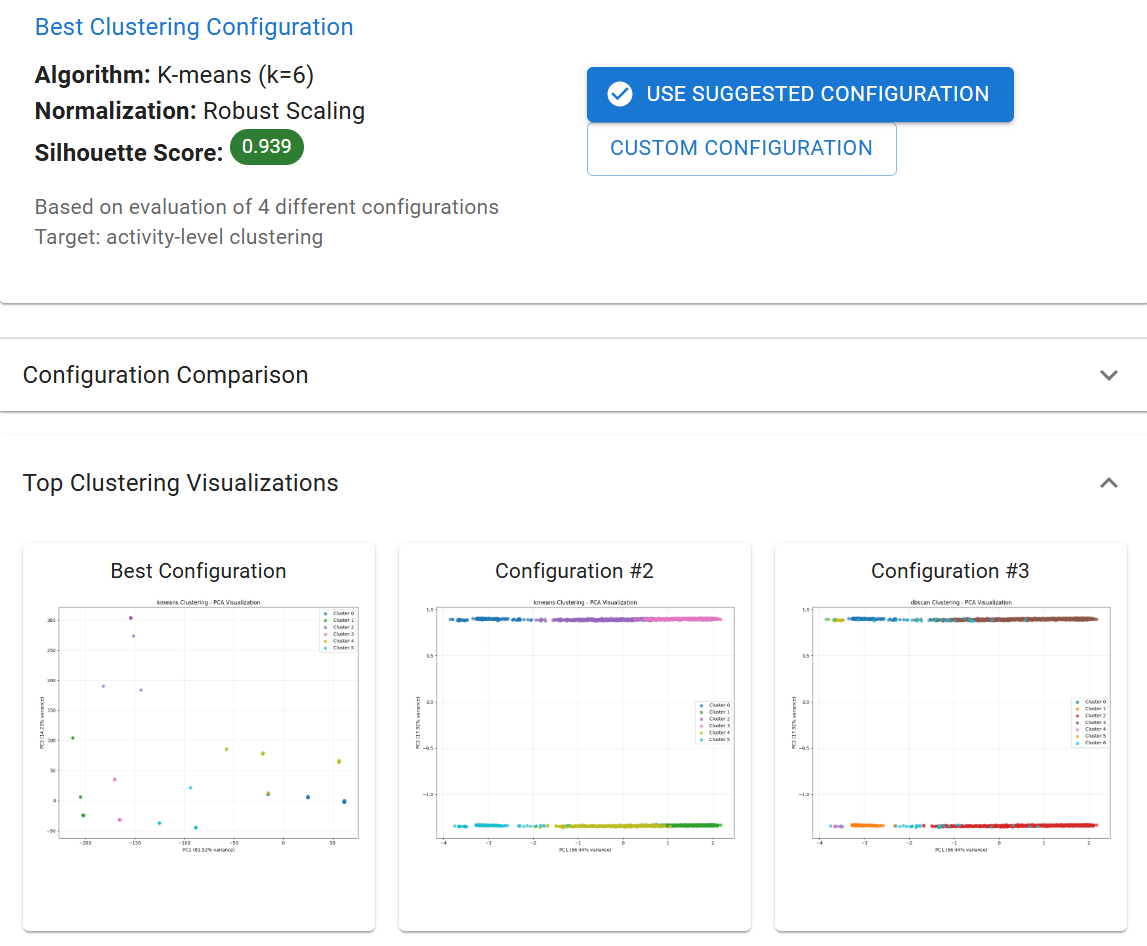}
\caption{Clustering results window}
\label{fig:clustering}
\end{figure}

Once the optimal clustering configuration is identified, IoT Miner generates detailed statistical profiles for each cluster, capturing the distinctive sensor characteristics that define different operational activities. These profiles provide a comprehensive characterization of each activity, including basic statistics like min, max, mean, median, and standard deviation, quartiles ($Q_1$, $Q_3$), helping to distinguish between different operational modes.   
To visualize high-dimensional cluster structures and validate their separation, we employ dimensionality reduction techniques (like Principal Component Analysis (PCA) and t-Distributed Stochastic Neighbor Embedding (t-SNE)).
These visualizations in both 2D and 3D help domain experts interpret the discovered patterns and validate the clustering results (Fig.~\ref{fig:clustering}). 

Unlike previous approaches that rely on predefined thresholds or fixed parameters, our method dynamically adapts to the statistical characteristics of each specific dataset. This adaptability is crucial for industrial settings where sensor data characteristics can vary significantly between different equipment, processes, and operational conditions.

\subsection{AI-Powered Labeling}

IoT Miner's key innovation is automated generation of semantically meaningful activity labels, bridging statistical cluster discovery with domain-appropriate terminology and eliminating manual interpretation bottlenecks.

\begin{figure}[htbp]
\centering
\begin{minipage}[b]{0.49\textwidth}
    \centering
    \includegraphics[width=\textwidth]{  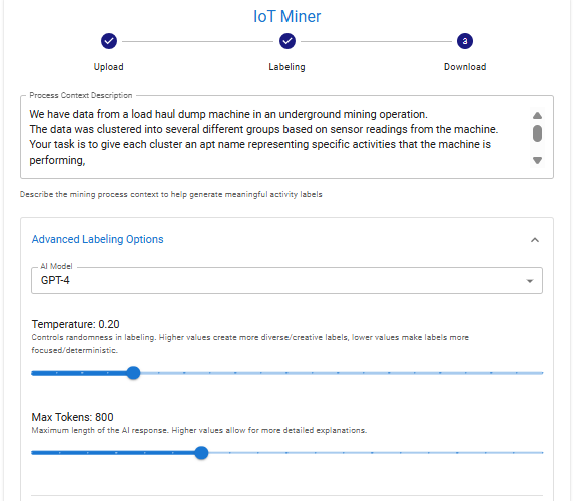}
    \caption{LLM labeling options window}
    \label{fig:labeling}
\end{minipage}
\hfill
\begin{minipage}[b]{0.49\textwidth}
    \centering
    \includegraphics[width=\textwidth]{  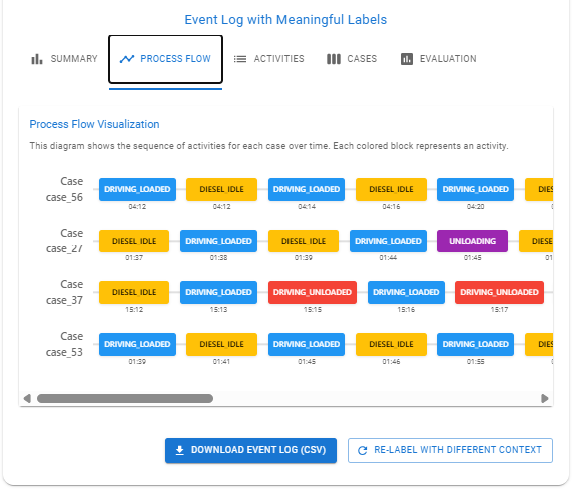}
    \caption{Event log construction interface showing fragment of created event log}
    \label{fig:eventlog}
\end{minipage}
\end{figure}

Generated statistical profiles of discovered clusters are automatically incorporated into the labeling prompt, which is extended with a user-provided context for the language model (Fig.~\ref{fig:labeling}).
The labeling window provides the following configuration options:
\begin{itemize}
\item Model Selection --- choice between different GPT models (GPT-4, GPT-3.5) based on complexity requirements,
\item Temperature Setting --- controls randomness in generation (0.0-1.0), with lower values producing more deterministic outputs,
\item Token Limits --- maximum response length configuration (typically 400-800 tokens) ensuring concise but complete labels,
\item Prompt Pool --- structured template for guiding the model toward appropriate industrial terminology.
\end{itemize}

The IoT Miner transmits configured prompts through Azure OpenAI API.
The API integration handles authentication, request formatting, and response processing while maintaining secure communication channels.

The language model processes combined statistical and contextual information to generate activity labels.
This automated labeling pipeline transforms abstract statistical clusters into interpretable activity descriptions, significantly reducing the expertise and time required for cluster interpretation.

\subsection{Event Log Construction}

The final stage transforms labeled data into event logs through structured case segmentation and temporal sequence construction, enabling standard process mining tool analysis.
This step allows configuring segmentation parameters, including the method of case detection (e.g., based on time gaps, days, or sensor changes), the sensitivity threshold for detecting significant changes, the minimum number of activities per case, and the maximum case duration. Additionally, consecutive identical activities can be optionally merged to reduce noise and improve event granularity.

The IoT Miner monitor key sensors for deviation patterns indicating operational transitions, identifies major activity transitions as natural boundaries, and applies configurable temporal constraints ensuring meaningful operational sequences. This addresses continuous industrial process challenges where traditional case boundaries lack a clear definition.
Constructed event logs reveal operational patterns previously invisible in raw sensor data, automatically identifying frequent activity transitions and enabling workflow discovery (Fig.~\ref{fig:eventlog}).
Generated event logs conform to standard XES format requirements (containing case ID, activity and timestamp), enabling direct import into established process mining tools (ProM, Celonis, Disco)~\cite{vanderAalst2022}. 

In the case of having the ground truth labels for dataset user can compare generated event logs against operational monitoring records, employing temporal matching and semantic similarity assessment.

\section{Evaluation}

\subsection{Dataset and Machine Description}

The dataset utilized in this study was collected from a room and pillar underground oil shale mine. It originates from a Load-Haul-Dump (LHD) machine, specifically the \textit{Scooptram ST1030} manufactured by \textit{Epiroc}. This LHD is designed for heavy-duty load and haul operations and features a nominal payload capacity of 10 tons. It is equipped with a diesel engine and an articulated steering mechanism optimized for confined underground conditions.
The Scooptram ST1030 operates alongside various other mining equipment, including drill rigs, roof bolters, and additional LHDs. Within this ecosystem, the LHD performs repeated loading, transporting, and dumping cycles under dynamically changing operational conditions. 

The dataset consists of raw signals captured via the machine's Controller Area Network (CAN) bus. The sensor streams offer a detailed view of the LHD’s mechanical and operational states. 
A sample with key recorded engine and transmission parameters is presented in Table~\ref{tab:sample}. These signals, such as engine speed (\textbf{ES}), torque (\textbf{TQ}), and fuel rate (\textbf{FR}), are crucial for assessing the machine’s operating conditions and performance patterns during mining operations.

\begin{table}[h!]
\centering
\begin{tabular}{|c|c|c|c|c|c|c|c|c|c|c|c|c|}
\hline
\textbf{Timestamp} & \textbf{~ES~~} & \textbf{~OP~} & \textbf{TQ} & \textbf{~FR~} & \textbf{APP} & \textbf{~DIT~} & \textbf{DOT} & \textbf{RS} & \textbf{MP} & \textbf{MT} & ~...~ \\
\hline
241001T06:03:02 & 712.875 & 456 & 35 & 9.15 & 0 & 25.809 & 5.535 & 0 & 10 & 11 & ~...~ \\
\hline
241001T06:03:03 & 725.875 & 456 & 36 & 6.6 & 0 & 26.181 & 5.566 & 0 & 10 & 11 & ~...~ \\
\hline
~... & ...~ & ...~ & ...~ & ...~ & ...~ & ...~ & ...~ & ...~ & ...~ & ...~ & ~...~ \\ \hline
\end{tabular}
\caption{Sample data collected from the LHD machine's engine and transmission system. Abbreviations: \textbf{ES} – Engine Speed (RPM), \textbf{OP} – Oil Pressure (kPa), \textbf{TQ} – Torque (\%), \textbf{FR} – Fuel Rate (L/h), \textbf{APP} – Accelerator Pedal Position (\%), \textbf{DIT} – DPF Intake Temp (°C), \textbf{DOT} – DPF Outlet Temp (°C), \textbf{RS} – Regeneration Status, \textbf{MP} – Manifold Pressure (kPa), \textbf{MT} – Manifold Temp (°C).}
\label{tab:sample}
\end{table}
These parameters collectively form a multi-dimensional time series that captures the low-level physical behavior of the machine during operation. 
As is typical for industrial sensor data, these recordings do not contain explicit information about tasks or activities, making them unusable for process mining in their raw form.
These signals exhibit, however, characteristic temporal and statistical patterns, which support the segmentation of machine behavior into distinct operational phases such as idling, loading, hauling, and dumping. As such, the dataset provides an effective benchmark for evaluating sensor-based event abstraction techniques under real-world industrial conditions.

\subsection{Experimental Setup}

To carry out our experiments, we first loaded time-series sensor data collected from an LHD machine. We focused our analysis on five key engine-related signals: accelerator pedal position, engine speed (RPM), engine oil pressure, fuel rate, and engine torque output. Based on these inputs, we applied clustering to discover recurring activity patterns. The optimal number of clusters was determined to be six, using K-means with Robust Scaling as the best-performing configuration (Silhouette Score: 0.939).
Next, the clustered data was segmented into activity cases using an automated sensor-based case detection method with the default IoT Miner parameters. 

To explore how prompt quality and contextual information affect label generation for clusters, we designed three experimental conditions with increasing levels of instruction and domain-specific guidance.

\begin{enumerate}
    \item Experiment 1 -- Basic Prompt without Task Clarification \\
    The model received a minimal prompt asking it to assign activity labels based on sensor statistics from clustered time-series data of an LHD machine. No~domain-specific context, examples, or constraints were provided (Fig.~\ref{fig:labeling}).

    \item Experiment 2 -- Prompt with Labeling Instructions\\
    This version extended the basic prompt by including a labeling instruction section that provided relevant activity examples and instructed the model to avoid ambiguous labels (e.g., those containing the word ``or'').

    \item Experiment 3 -- Prompt with Context and Operational Guidance \\
    In addition to the improvements from Experiment 2, this prompt included detailed background information about the LHD machine's operational cycle and interpretive guidelines for reading sensor patterns. It emphasized specific mappings between sensor values and mining-related activities and encouraged unambiguous, domain-appropriate labeling (see Prompt Box~\ref{box:questionprompt1}).
\end{enumerate}

\setcounter{promptboxcounter}{1}
\begin{promptbox}[box:questionprompt1]{Prompt for Experiment 3}
We have time-series sensor data collected from a Load-Haul-Dump (LHD) machine operating in an underground mining environment. The machine follows a repetitive operational cycle consisting of distinct phases: loading blasted rock, hauling it through tunnels, dumping it, and returning to the loading point. 
The machine may also spend time idling or stopped between activities.

This data has been clustered into several distinct groups based on patterns in the following engine-related sensor values:

- engine 1 accelerator pedal position 1: Degree to which the operator is pressing the accelerator pedal (0–100\%)

- engine 1 engine speed: Revolutions per minute (RPM) of the engine

\textit{[here more description of sensor values]}

Use mining-related terminology to describe each operational context 



\textit{[here more description of mining-related terminology]}



Consider the following operational insights when naming the clusters:
Engine speed (RPM) alone does not distinguish phases. 
Accelerator pedal position indicates operator intent or system command. A high value suggests active tramming or bucket work; low values suggest coast, idle, or no operator input.
Fuel rate and torque increase under higher load demands, such as lifting, tramming uphill, or carrying heavy material.

\textit{[here more operational description of the task]}

\end{promptbox}

For each experiment, we prompted the GPT-4 model using six different temperature values (ranging from 0.0 to 1.0 in steps of 0.2), generating 10 labeled event logs per temperature to enable consistent comparison.

In our experiments, due semantic variations between the predicted and ground truth activity labels, a direct one-to-one label matching was not feasible. 
To address the challenge of non-aligned labels between predicted and reference activity names, we introduce a metric that accounts for their semantic similarity. Instead of relying on exact string matches, this approach assigns partial credit based on how closely the labels relate in meaning. We refer to this metric as Similarity-Weighted Accuracy (SWA), and define it formally as follows:

\[
\text{SWA} = \frac{1}{N} \sum_{i=1}^{N} \left( \mathbb{1}_{[s_i \geq T]} \cdot s_i \right),
\]

\noindent where:
\begin{itemize}
  \item $N$ is the total number of reference instances,
  \item $s_i \in [0, 1]$ is the similarity score between the predicted and reference label for instance $i$,
  \item $T \in [0, 1]$ is the similarity threshold,
  \item $\mathbb{1}_{[s_i \geq T]}$ is the indicator function: it equals 1 if $s_i \geq T$, and 0 otherwise.
\end{itemize}

To measure the similarity score $s_i$ between the predicted and reference labels for each instance, we used Google's \texttt{models/embedding-001} text embedding model.
This formulation discards similarity scores below the threshold and retains the full value of those above it, allowing partial credit only for sufficiently similar predictions.

\begin{figure}[htbp]
    \centering
    \includegraphics[width=.99\textwidth]{  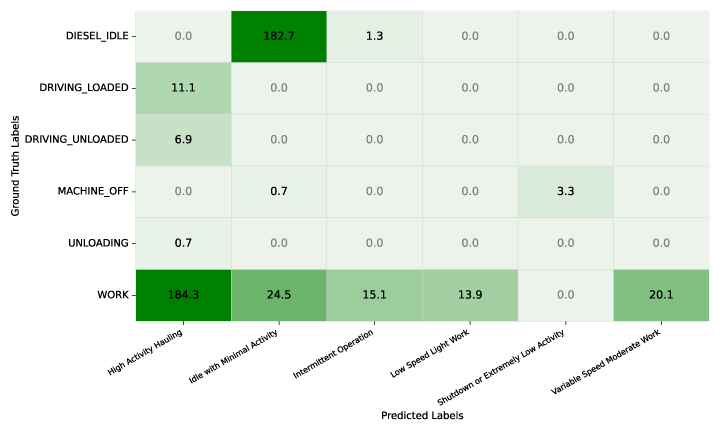}
    \caption{Similarity-weighted accuracy heatmap showing the correspondence between predicted and ground truth labels.}
    \label{fig:similarity_heatmap}
\end{figure}

Fig.~\ref{fig:similarity_heatmap} presents an example of a similarity-weighted accuracy matrix used to evaluate the alignment between predicted and ground truth labels. Since the predicted labels come from unsupervised clustering, they are not guaranteed to align one-to-one with the actual activity classes—multiple ground truth classes can be grouped into a single cluster, or a single class may be split across several clusters. Additionally, the event distribution in the log is imbalanced, which may also affect the density of values in the matrix.
The resulting matrix reflects the product of semantic similarity scores and the corresponding frequencies of label co-occurrence, highlighting meaningful overlaps between predicted and ground truth classes (the color intensity indicates regions of higher weighted agreement).

\subsection{Results}

To ensure robustness of the evaluation and account for prediction variability, each experiment–temperature configuration was tested in 10 independent runs. 

\begin{figure}[hb!]    
    \centering
    \includegraphics[width=0.95\textwidth]{  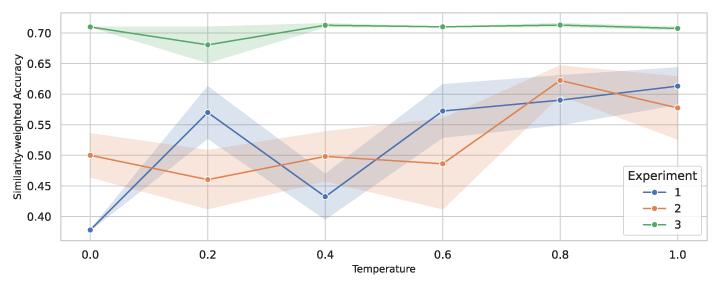}
    \caption{Average similarity-weighted accuracy by temperature per experiment.}
    \label{fig:avg-accuracy-lineplot}
\end{figure}

The resulting performance trends are visualized in Fig.~\ref{fig:avg-accuracy-lineplot}, which shows the average similarity-weighted accuracy across different temperature levels for each experiment. The shaded areas around the lines represent the standard error of the mean (SE), reflecting how much the average score might vary due to randomness across runs. Narrower bands suggest more consistent and reliable performance under that condition.

\begin{figure}[htbp]
    \centering
    \includegraphics[width=.99\textwidth]{  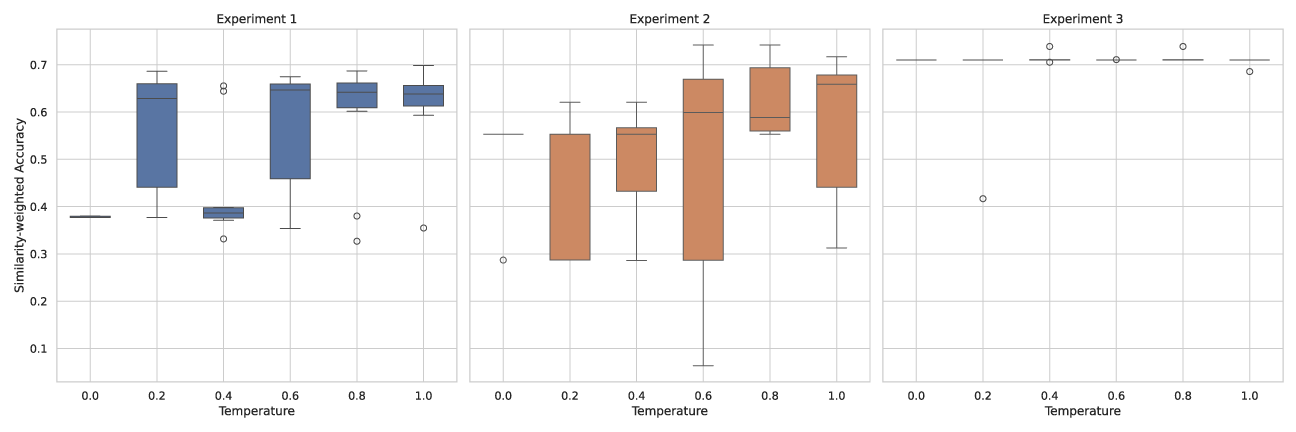}
    \caption{Distribution of similarity-weighted accuracy for each experiment across different temperature settings.}
    \label{fig:boxplots-experiments}
\end{figure}

To further explore variability and stability, Fig.~\ref{fig:boxplots-experiments} presents a distribution of the similarity-weighted accuracy across the three separate experiments. Each boxplot shows the accuracy distributions for various temperature settings presenting variability and performance consistency within and between experiments.

\begin{figure}[t!]
    \centering
    \includegraphics[width=.78\textwidth]{  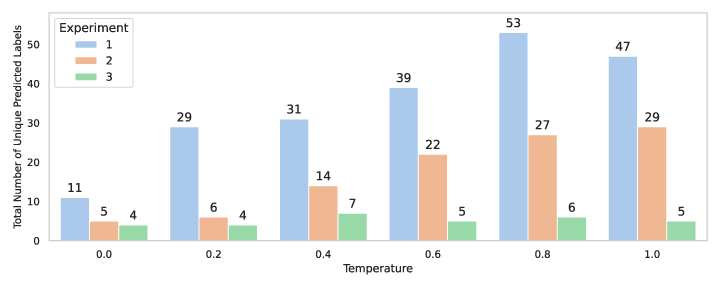}
    \caption{Total number of unique predicted activity labels in experiments per temp.}
    \label{fig:unique-labels-barplot}
\end{figure}

Beyond accuracy alone, we also examined the diversity of predicted activity labels across conditions. As shown in Fig.~\ref{fig:unique-labels-barplot}, the number of distinct labels (aggregated over 10 runs per temperature) varies notably with both temperature and prompt quality. 
The results show that the basic prompt (Experiment~1) consistently produces a higher number of unique labels, particularly at higher temperatures, reflecting a more diverse and unconstrained interpretation of cluster semantics. In contrast, Experiment 3, which includes detailed operational context and labeling guidance, tends to result in fewer and more consistent labels—indicating that domain-specific instructions may help the model converge toward a narrower, more uniform naming scheme.

\subsection{Discussion and limitations}

Our approach assumes that the input IoT data consists mainly of numerical or boolean time-series data (e.g., RPM, fuel rate, pressure), as commonly found in industrial systems. This makes it well-suited for industrial sensor streams, but less applicable to other data types like images, audio, or complex inputs. This limits the generalizability of the approach to environments where sensor signals show temporal and statistical patterns that can be grouped using clustering.

The approach uses ML methods such as clustering and LLMs, which are generalizable and can be applied to different datasets. However, the formulation of effective prompts remains a critical factor for label quality. In our setup, prompt creation is left to the user, including how to describe the sensors, explain the machine's operation, and guide the label generation process. While our tests show that well-written prompts improve performance, the current version does not provide built-in help for this step.

Our pipeline includes configurable parameters at multiple stages. Although the default settings often produce satisfactory results, selecting appropriate parameters or case segmentation settings may still require expert input and can be challenging for non-expert users without system support. Moreover, our evaluation focused on a single, yet real-world mining use case. This domain offers a rich source of sensor data, it is also highly specific and complex, with limited representation in general LLM training data. As a result, generating accurate and meaningful labels can be more difficult compared to more common and well-documented domains. The broader applicability of the approach in other industrial contexts still needs to be validated.

In the case of evaluation-related limitations, the ground truth labels may include uncertainty or simplification, especially in industrial settings where activity boundaries are fuzzy (often unclear or overlapping). This makes it difficult to assign a single correct label, which in turn affects the accuracy of comparisons between predicted and reference labels. Additionally, our results rely on specific versions of LLMs accessed via external APIs. Any changes to model behavior, availability, or access policies could impact the reproducibility and consistency of the results. Even when provided with clear instructions, LLMs may still generate inconsistent or misleading labels, especially at higher temperature settings or when prompts lack sufficient context. This may raise concerns about the reliability of the output. To address these challenges, more support in writing prompts, testing in other domains, and guiding users are recommended.

\section{Conclusion and Future Work}

In this paper, we introduced IoT Miner, an unsupervised framework that transforms raw industrial sensor data into high-level event logs suitable for process mining. Our approach combines domain-aware preprocessing, comparative clustering, and semantic labeling powered by LLMs. This enables automatic discovery and interpretation of operational activities from raw sensor streams without requiring expert-defined rules or labeled data.

The experimental evaluation on real-world data from an LHD mining machine demonstrated the effectiveness of IoT Miner. The automatically discovered best clustering configuration produced highly cohesive activity clusters, and LLM labeling showed that more detailed prompts lead to more accurate and consistent labels, highlighting the importance of context when using LLMs in industry.

With flexible configuration options at various stages, IoT Miner adapts to diverse industrial settings—from manufacturing to energy production—without extensive manual setup.
Overall, our approach automates transformation of high-frequency industrial sensor data into semantically meaningful process representations, addressing critical gaps in industrial process mining while maintaining domain interpretability and tool compatibility.

In future work, we plan to integrate additional data sources—such as images or operator logs—to improve clustering and labeling accuracy. We also aim to benchmark IoT Miner across a range of industrial datasets to evaluate its generalizability. Furthermore, optimizing the pipeline for real-time and edge use would enable fast, on-site monitoring, while incorporating optional expert feedback could support label verification, combining automation with human insight.


\end{document}